\documentclass[conference]{IEEEtran}
\IEEEoverridecommandlockouts
\usepackage{cite}
\usepackage{amsmath,amssymb,amsfonts}
\usepackage{algorithmic}
\usepackage{graphicx}
\usepackage{textcomp}
\usepackage{balance}
\usepackage{tikz,tkz-tab}%package pour les tableaux de variations et de signes%
\usepackage{circuitikz}
\usepackage{changes}
\definechangesauthor[color=blue,name={Thomas Luinaud}]{TL}
\definechangesauthor[color=orange,name={Thibaut Stimpfling}]{TS}
\definechangesauthor[color=red,name={Jeferson Santiago Da Silva}]{JS}
\usetikzlibrary{positioning, automata, graphs, trees, fit, arrows.meta, shapes}
\usetikzlibrary{backgrounds,patterns,matrix,calc,shadows,plotmarks, circuits.logic.US}
\usetikzlibrary{decorations}
\usepackage[detect-weight=true, load-configurations=binary]{siunitx}
\DeclareSIUnit{\bit}{b}
\usepackage{xcolor}
\def\BibTeX{{\rm B\kern-.05em{\sc i\kern-.025em b}\kern-.08em
    T\kern-.1667em\lower.7ex\hbox{E}\kern-.125emX}}
    
\begin{document}
\title{One for All, All for One: A Heterogeneous Data Plane for Flexible P4 Processing}
\author{
	\IEEEauthorblockN{Jeferson Santiago da Silva\textsuperscript{1}, Thibaut Stimpfling\textsuperscript{1}, Thomas Luinaud\textsuperscript{1}, Bachir Fradj\textsuperscript{1} and Bochra Boughzala\textsuperscript{2}}   
	\IEEEauthorblockA{\textsuperscript{1}Polytechnique Montr\'{e}al, Canada, \{firstname.lastname\}@polymtl.ca, \textsuperscript{2}Kaloom Inc., Montr\'{e}al, Canada, bochra@kaloom.com}
	\thanks{The first four authors have equally contributed to this work.}
	\thanks{This work is supported by CNPq/Brazil, Mitacs/Canada, and Kaloom Inc.}
%	\IEEEauthorblockN{Hidden for Blind Review}
}
\maketitle

\begin{abstract}
The P4 community has recently put significant effort to increase the diversity of targets on which P4 programs can be implemented. These include fixed function and programmable ASICs, FPGAs, NICs, and CPUs. However, P4 programs are written according to the set of functionalities supported by the target for which they are compiled.
For instance, a P4 program targeting a programmable ASIC cannot be extended with user-defined processing modules, which limits the flexibility and the abstraction of P4 programs.

To address these shortcomings, we propose a heterogeneous P4 programmable data plane comprised of different targets that together appear as a single logical unit. The proposed data plane broadens the range of functionalities available to P4 programmers by combining the strength of each target. We demonstrate the feasibility of the proposed P4 data plane by coupling an FPGA with a soft switch which emulates a programmable ASIC. The proposed data plane is demonstrated with the implementation of a simplified L2 switch.
The emulated ASIC match-table capacity is extended by the FPGA by an order of magnitude.The FPGA also integrates a proprietary module using a P4 extern.
\end{abstract}


\section{Introduction}
Data center applications have pushed the envelope towards high-throughput yet programmable network devices. Such a trade-off can be alleviated by recent advancements in programmable ASICs and network programming languages as P4 \cite{Bosshart:14}. However, to achieve higher throughput, these programmable ASICs limit the functionalities that can be expressed in P4 due to tighter hardware constraints.
% In number ? 
For instance, integrating new proprietary modules with P4 externs is not supported on a programmable ASIC. 

In addition, standalone hardware platforms supporting P4, shown in Table~\ref{tab:relation_tagets}, exhibit an inherent trade-off between the degree of programmability and performance. To overcome these limitations, complementary P4 targets are required.
\begin{table}[ht]
	\centering
	\caption{Available P4 targets}
	\label{tab:relation_tagets}
	\begin{tabular}{lllll}
		\hline
		Target &  \begin{tabular}[c]{@{}l@{}}Degree of\\Programmability\end{tabular} & Throughput & Latency      & \begin{tabular}[c]{@{}l@{}}Power\\Efficiency\end{tabular} \\
		\hline
		ASIC   & Limited         & Very high  & Very low     & Very High        \\
		FPGA   & High            & High       & Low/very low & High             \\
		NIC    & High            & Limited    & Limited     & High             \\
		%GPU    & Very high       & Limited*   & Very high*   & Very low         \\
		CPU    & Very high       & Very low   & Very high    & Low              \\
		\hline
	\end{tabular}
\end{table}

While some research has been conducted to diversify the number of P4 targets \cite{Dang:2017}, little work as been devoted in exploiting the strength of complementary targets. We believe that combining multiple targets into a heterogeneous hardware platform allows filling this gap.

% Peut etre expliquer interet de plat heteorgenous des le debut.
In this paper, we introduce the concept of a heterogeneous data plane (HDP) platform. As a proof of concept, we present a HDP comprising an ASIC and an FPGA. By coupling a programmable ASIC and an FPGA, we extend the set of functionalities that can be described using P4. The FPGA programmability enables the implementation of custom P4 pipelines, which covers proprietary modules expressed as P4 externs, while increasing hardware resources. 

In contrast to previous works that have used FPGAs only as hardware accelerators \cite{Caulfield:2018}, the FPGA is here integrated as a main component of a logical P4 pipeline. 

The main contribution of this work is to start a discussion towards heterogeneous data-plane platform. We present a proof of concept that:
\begin{itemize}
	\item  Combines multiple targets into a logical P4 pipeline.
	\item  Exploits the programmability characteristics of FPGAs to extend the limitations of an ASIC.
	\item  Implements a proprietary module as a P4 extern.
	 
\end{itemize}

We believe that HDP will open new doors to more innovative P4 programs and network applications.

\section{Heterogeneous Data-Plane}

%Referring to Table~\ref{tab:relation_tagets}, one would like to achieve the performance of a programmable ASIC while keeping the programmability flexibility of a CPU. However, this cannot be achieved using a single target P4 pipeline. A feasible solution lies in using a heterogeneous platform to extract the strengths of each individual target.

A heterogeneous data plane (HDP) comprises different P4 targets. Loosely coupling diverse targets extends the range of functionalities and applications that can be expressed in P4.
%Yet, this allows the creation of an elastic P4 pipeline. By \textbf{elastic}, we mean that P4 applications can be \textbf{elastically} extended over the targets, how long there are enough hardware resources to do so.
%Yet, this allows the creation of an adaptable pipeline that can be extended over the targets.% This extension is only limited by the available hardware resources.

A generic view of a HDP is depicted in Fig.~\ref{fig:hetero_overview}. In this figure, multiple P4 targets are connected together to form a single logical P4 pipeline. Similarly, a control-plane entity manages the HDP. A compiler tool chain splits the original P4 description across the different targets. %, exploiting the forces of each one.
The proposed architecture is described in the following subsections.

\subsection{Hardware and Communication Infrastructure}

The proposed HDP is composed of multiple P4 targets abstracted into a single logical P4 target. Hence, a HDP can combine any set of P4 targets, from x86 servers, GPUs, NICs, to FPGAs and ASICs. %%%%%%%%%%%%%%%%%%%%%%%%%%%%%%%%%%%%%%%%%%%%%%%%%%%%%%%%%%%%%%%%%%%%%%%%%%%%%%%%%%%%%%%%%%%%%%%%%%%%%%%%% However, targets should be selected carefully to extend the programmability capabilities, while balancing the performance. % Non claire ce que performance veut dire ici. la p 
One HDP example is presented in the rounded rectangle of Figure~\ref{fig:hetero_overview}. %Decrire la figure de Thomas.

The communication links between the P4 targets of a HDP does not require any specific protocol nor medium. 
Since the proposed HDP is programmed with P4, there is no communication protocol requirements between targets. The only constraint lies in the feasibility of describing the chosen protocol using P4 idioms. Thus, to connect multiple targets in a single HDP, different media may be used, ranging from PCIe, Ethernet-capable links, QPI, to proprietary communication interfaces. For instance, GPUs, or FPGAs connected to a server are likely to use a PCIe link, while an ASIC connected to an FPGA could use an Ethernet link. 

\subsection{Compilation Process}

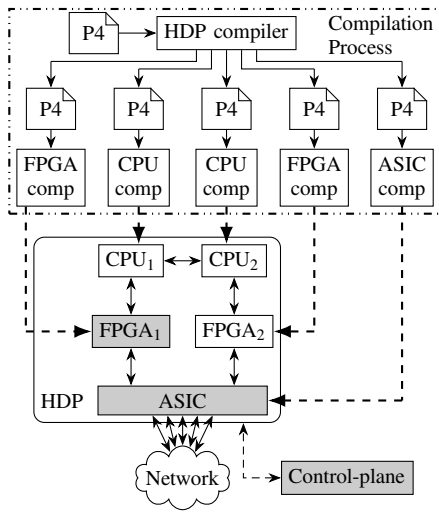
\begin{figure}[t]
	\centering
%	\centerline{\includegraphics{fig1.png}}
	\begin{tikzpicture}[node distance = 0.5cm, every node/.style={draw}, node font={\footnotesize}]
%software
\begin{scope}[every node/.style={draw, chamfered rectangle, chamfered rectangle corners=north east}]
	\node (globp4) {P4};
	\node[rectangle, right=of globp4] (compiler) {HDP compiler};
	\node[below= 0.5cm of compiler] (p4cpu2) { P4};
	\node[left= of p4cpu2] (p4cpu1) {P4};
	\node[left= of p4cpu1] (p4fpga1) {P4};
	\node[right= of p4cpu2] (p4fpga2) {P4};
	\node[right= of p4fpga2] (p4asic) { P4};
%%path for the corner of p4 codes
	\path[draw]  (globp4.before north east) -| (globp4.after north east)
							  (p4cpu2.before north east) -| (p4cpu2.after north east)
							  (p4cpu1.before north east) -| (p4cpu1.after north east)
						      (p4fpga1.before north east) -| (p4fpga1.after north east)
							  (p4fpga2.before north east) -| (p4fpga2.after north east)
							  (p4asic.before north east) -| (p4asic.after north east);

%%compilation
	\begin{scope}[every node/.style={rectangle, draw, align=center}, node distance=0.25cm]
		\node[below= of p4cpu2] (compcpu2) { CPU\\ comp};
		\node[below= of p4cpu1] (compcpu1) { CPU\\ comp};
		\node[below= of p4fpga1] (compfpga1) { FPGA\\ comp};
		\node[below= of p4fpga2] (compfpga2) { FPGA\\ comp};
		\node[below= of p4asic] (compasic) { ASIC\\ comp};
	\end{scope}
%%processus
	\begin{scope}[every edge/.style={-Stealth, draw}]
   \coordinate (midBelowComp) at ($(compiler.south) - (0,0.25)$);
	\path[draw] (globp4) edge (compiler)
	    							  (compiler.south) edge (p4cpu2)
								  ($(compiler.south) - (0.2,0)$) |- ( p4cpu1.north |- midBelowComp) edge (p4cpu1)
                                  ($(compiler.south) - (0.4,0)$) |- ($( p4fpga1.north |- midBelowComp) + (0,0.1)$) edge (p4fpga1)
								 ($(compiler.south) + (0.2,0)$) |- ( p4fpga2.north |- midBelowComp) edge (p4fpga2)
                                  ($(compiler.south) + (0.4,0)$) |- ($( p4asic.north |- midBelowComp) + (0,0.1)$) edge (p4asic)
								 (p4cpu1) edge (compcpu1)
								 (p4cpu2) edge (compcpu2)
								 (p4fpga1) edge (compfpga1)
								 (p4fpga2) edge (compfpga2)
								 (p4asic) edge (compasic)
									;
	\end{scope}

\end{scope}
\coordinate (midCPU) at ($(p4cpu1.east |- compcpu1.south) + (0.25cm, -0.5cm)$);
%%hardware
\begin{scope}[every node/.style={draw},]
	\node[left=0.25cm of midCPU, anchor=north east] (cpu1) {CPU\textsubscript{1}};
	\node[right=0.25cm of midCPU, anchor=north west] (cpu2) {CPU\textsubscript{2}};
	\node[below=of cpu1, fill=black!20] (fpga1) {FPGA\textsubscript{1}};
	\node[below=of cpu2] (fpga2) {FPGA\textsubscript{2}};
	\node[fill=black!20, minimum width=2.25cm, anchor = north] (asic) at ($(midCPU|-fpga2.south) - (0,0.5)$) {ASIC};
	\node[draw=none,left =0.55 of asic, align=left] (data_label) {};
	
\end{scope}
\node[fit=(cpu1) (cpu2) (fpga1) (fpga2) (asic) (data_label), rounded corners] (platform) {};
\node[draw=none, anchor=west] at (platform.west |- asic.east) {HDP};
\node[cloud, draw, aspect=2, inner sep=-2pt, below=0.4cm of asic] (network) {Network};
\begin{scope}[every edge/.style={Stealth-Stealth, draw}]
\path (cpu1) edge (cpu2)
			(cpu1) edge (fpga1)
			(cpu2) edge (fpga2)
			(fpga1.south) edge (fpga1.south |- asic.north)
			(fpga2.south) edge (fpga2.south |- asic.north)
			;
\end{scope}
\begin{scope}[every edge/.style={thick, -Latex, dashed, draw}, every path/.style={thick, dashed}]
		\path[draw] ($(platform.west |- compfpga1.south) - (0.1,0)$) -- ($(platform.west |- fpga1) - (0.1,0)$)  edge (fpga1)
								(compfpga2) -- (compfpga2 |- fpga2) edge (fpga2)
								(compasic) -- (compasic |- asic) edge (asic)
								(compcpu2) edge (compcpu2.south |- cpu2.north) 
								(compcpu1) edge (compcpu1.south |- cpu1.north) 
								;
	\end{scope}
\path[Stealth-Stealth, draw] (network) edge (asic.south)
 							   (network) edge ($(asic.south) + (0.2,0cm)$)
 							   (network) edge ($(asic.south) - (0.2,0cm)$)
 							   (network) edge ($(asic.south) + (0.4,0cm)$)
 							   (network) edge ($(asic.south) - (0.4,0cm)$);
	\node[draw, anchor = west, fill=black!20](control) at (network.east -| platform.east) { Control-plane};	
	  \path[densely dashed, Stealth-Stealth, draw] (control) -| ($(platform.south east) - (0.5cm,0)$);
	\node[fit = (compfpga1) (compasic) (compiler), draw, dash dot dot, thick] (comp) {};
	\node[draw=none, below left=0 and 0 of comp.north east, align=left] { Compilation\\ Process};
	\end{tikzpicture}
	
	
	\caption{Overview of a Heterogeneous Data-Plane (HDP)}
	\label{fig:hetero_overview}
\end{figure}

As shown in the dashed rectangle of Fig.~\ref{fig:hetero_overview}, the HDP compiler splits a P4 program into multiples target-specific P4 programs. Each generated P4 program implements only a portion of the functionality expressed in the original P4 program. The challenge of HDP compiler is to efficiently partition the original P4 program and to map each functionality to a specific target, based on performance constraints and user needs given in a configuration file.%New research avenues should cover partitioning mechanisms to efficiently schedule a P4 program on multiple targets, based on resources availability and user needs.

\section{Proof of concept}

As a proof of concept, we propose a HDP highlighted in grey boxes of Fig.~\ref{fig:hetero_overview}. The proposed HDP contains an ASIC that only includes basic P4 functionalities and an FPGA. On this HDP, we have implemented a L2 switch with a packet counter. The HDP compiler is not covered in this work as we have manually mapped the P4 program to each target.

%As a proof of concept, we have implemented a L2 switch with a packet counter on the heterogeneous data-plane highlighted in grey boxes in Fig.~\ref{fig:hetero_overview}. The HDP comprises an ASIC that only includes basic P4 functionalities and an FPGA. The HDP compiler is not covered in this work as we have manually mapped the P4 program to each target.
%The HDP comprises an ASIC that only includes basic P4 functionalities and an FPGA.

The L2 switch is divided into the ASIC and the FPGA. Hence, the L2 MAT entries are divided between the ASIC and the FPGA.
In addition, the packet counter, implemented on the FPGA as a P4 extern, computes the number of packets matched on the FPGA. Thus, the FPGA extends an unsupported ASIC feature. In addition, an external DDR3 DIMM connected to the FPGA increases by an order of magnitude the size of the L2 MAT compared to an ASIC. 

%However, the behavior described by the two P4 programs  The P4 program executed on the ASIC However, the two P4 program describe a single data flow such as  executed on the FPGA p r Thus the overhead of the proposed heterogeneous data-plane is limited as the P4 program is not duplicated for both targets. 

% Comment dans la demo on extend la memoire du ASIC sur le FPGA ? Selon quelle critere ?

% Quel filtre ?

%Two types of configuration ASIC -> FPGA or ASIC-> FPGA -> ASIC

%AJOUTER UN SCHEMA AVEC UNE REPRESENTATION LOGIQUE DU PIPELINE P4.

%TODO: use an abreviation for heterogeneous data-plane (HDP)

Because we had no access to a programmable ASIC, the bmv2 soft switch was used as an ASIC emulator. The bmv2 runs on a \texttt{x86} server equipped with a \SI{10}{\giga\bit/\second} Intel NIC. The FPGA P4 pipeline is compiled by Xilinx SDNet 2017.4, synthesized, and implemented using Xilinx Vivado 2018.1 on a Xilinx ZC706 evaluation board. The bmv2 and FPGA are connected using an Ethernet link. A simplified controller runs in an \texttt{x86} computer and communicates with the proposed architecture using a generic API.
Qualitative results achieved by the proposed HDP are presented in Table~\ref{tab:eval}.

\begin{table}[t]
	\centering
	\caption{Qualitative platform evaluation}
	\label{tab:eval}
	\begin{tabular}{llll}
		\hline
		Platform       & \begin{tabular}[c]{@{}l@{}}Generic\\externs\end{tabular} & \begin{tabular}[c]{@{}l@{}}Extensible\\M-A Tables\end{tabular} & \begin{tabular}[c]{@{}l@{}}Line-Rate\\Processing\end{tabular}  \\
		\hline
		FPGA           & $\surd$           & $\surd$                    & $\times$                    \\
		ASIC           & $\times$          & $\times$                   & $\surd$                    \\
		Proposed work  & $\surd$           & $\surd$                    & $\surd$                    \\
		\hline
	\end{tabular}
\end{table}

\section{Discussion and Research Directions}

% To date, the FPGA augments the ASIC, the contrary is true?
% In addition, both the on-chip FPGA resources of the FPGA and the augmented FPGA I/O capabilities can be extend the ASIC functionalities with dedicated co-processors, increasing thus the pool of resources available for a P4 program. 

% For instance, an external memory connected to the FPGA can be used to extend the number of entries inserted in match-action tables. In this work, we have used an off-the-shelf DDR3 increasing match-tables capacity by a factor of A CARACTERISER (CHIFFRES)
This section presents the challenges and research directions related to heterogeneous data planes. 

One challenge faced by a HDP relates to mismatched target performances. For instance, a lookup engine using an external memory connected to an FPGA may not match the lookup rate of an ASIC. Thus, the throughput of a HDP could be limited by a single target. However, in the proposed proof of concept, this limitation can be alleviated by using caching strategies. That is, the external memory connected to the FPGA would be seen as the main memory, while the ASIC memory would be used as a cache. Another solution can employ multiple parallel FPGA pipelines.

%	Coupling to task/HW manager, P4 applications can be scheduled to specific HW blocks, based on resources availability and user needs. In addition, an exotic application would be to use this architecture as in-network computation engine in case of little network activity. Such engines can exposed to users as HaaS elements using user-defined P4 externs.

In addition, a HDP introduces resources overhead. 
One resource overhead relates to the replication of parsers and deparsers on the HDP targets.
Finally, the HDP can introduce a variable latency as a packet may traverse multiple targets. 
However, these impacts may be reduced by using custom protocols to communicate between the HDP targets.

%First, parsers and desparsers are replicated in each target. However, custom protocols can be used to reduce resource waste. Moreover, multiple parallel FPGA pipelines may be required to match the throughput of state-of-the-art programmable ASIC, which  increases the resource consumption. Finally, the HDP can introduce a variable latency as a packet may traverse multiple targets.

% DIScussion challenge research / sur le compilateur. 
%Increase latency (carateriser). Third ?
Still, we believe the new ideas allowed by a HDP outweigh its overhead. As future work, we plan to develop the compiler tool-chain, which is the key element to leverage the HDP. 
Lastly, we hope that HDP ideas will be assessed and challenged by P4 enthusiasts to refine the proposed platform.
% Parler du surcout. 

\section*{Acknowledgment}
We would like to thank the anonymous reviewers for their comments and our shepherd Prof. Pierre Langlois. 
%Hidden for blind review.

\bibliographystyle{IEEEtran}
\bibliography{IEEEabrv,biblio}

\balance

\end{document}